\documentclass[a4paper,fleqn]{cas-sc}

\usepackage[numbers, sort&compress]{natbib}
\usepackage{subfigure}
\usepackage{multirow}
\usepackage{tabularx}
\usepackage{placeins} 
\usepackage{lineno}

\begin{document}


\title [mode = title]{Development of neutron transmutation doped germanium (NTD-Ge) for cryogenic applications}



%
\author[1]{Kangkang Zhao}
\fnmark[1]
\fntext[1]{Now at Gran Sasso Science Institute, Italy}
\author[1]{Mingxuan Xue}

\cormark[1]
\cortext[1]{Corresponding author}
\author[1]{Haiping Peng}
\cormark[1]
\cortext[1]{Corresponding author}
\author[1]{Deyong Duan}
\author[1]{Yunlong Zhang}
\author[1]{Yi Li}
\author[1]{Junfeng Yang}
\author[1]{Xintan Deng}
\author[1]{Hongjun Zhang}
\author[2]{Huaichang Ran}
\author[1]{Sicheng Wen}
\author[1]{Xiaolian Wang}
\author[1]{Zizong Xu}
\affiliation[1]{organization={Department of Modern Physics, University of Science and Technology of China},
            addressline={No.96 Jinzhai Road}, 
            city={Hefei},
            postcode={230026}, 
            state={Anhui},
            country={China}}
\affiliation[2]{organization={China Institute of Atomic Energy},
            city={Beijing},
            postcode={102413}, 
            country={China}}










\begin{abstract}
This paper presents the systematic fabrication and characterization of cryogenic thermometers based on neutron transmutation doped germanium (NTD-Ge).
High-purity (10N) germanium samples were irradiated by thermal neutrons with different fluences at the China Advanced Research Reactor (CARR). 
After irradiation and a $\sim$five-month cooling-down period, positron annihilation lifetime spectroscopy and temperature-dependent Hall effect measurements were performed to characterize irradiation-induced defects and carrier concentrations in the NTD-Ge samples. 
Utilizing standard semiconductor fabrication techniques, point electrodes were deposited onto the processed samples to fabricate functional NTD-Ge cryogenic thermometers. 
The low-temperature resistance performance of the devices was characterized down to 20~mK on a millikelvin range cryogenic test platform. 
The measured temperature dependence of resistance follows Mott's law, showing excellent agreement across the full measured range. 
The extracted $T_0$ is consistent with expectations. 
These results collectively verified both applicability of the thermometers in cryogenic system and reliability of the fabrication procedure.
\end{abstract}


\maketitle
\section{Introduction}\label{sec::intro}

With compelling fundamental research efforts in nuclear and astro-particle physics, cryogenic detectors equipped with diverse thermometry techniques for signal readout have witnessed rapid progress over recent decades~\cite{Enss:2005md, AMoRE:2024tjb, Pavan:2025zei, angloher:2026lhq, Quaranta:2026lin}. 
The key performance metrics of cryogenic thermometers, sensitivity, dynamic temperature range, time response characteristics and multiplexing readout capability, are essential for the practical deployment of such detectors. 
Among available cryogenic thermometer options, resistive neutron transmutation doped germanium (NTD-Ge) thermometers exhibit distinctive merits, namely their straightforward readout architecture, broad dynamic operating range and outstanding energy resolution, particularly for MeV-scale energy measurements~\cite{Enss:2005md, Haller1996NeutronTD, 2023APSHAWE12003D, Navick:2016ect, Mathimalar:2014sfa}. Additionally, NTD-Ge thermometers are also applicable to high-precision, wide-range temperature monitoring in industrial scenarios.

NTD-Ge refers to doped germanium semiconductors manufactured using the neutron transmutation doping (NTD) technique. 
Through thermal neutron capture inside a nuclear reactor and subsequent radioactive decay, the NTD method achieves excellent homogeneity of doping distribution within semiconductor materials. 
NTD-Ge with optimized doping concentrations possesses a pronounced temperature-dependent resistance effect, especially at ultra-low temperatures down to approximately 10 millikelvin (mK). 
Furthermore, its doping level can be precisely controlled, rendering NTD-Ge a stable and highly reproducible temperature sensor amenable to large-scale fabrication.

During the neutron irradiation process of germanium, different germanium isotopes capture thermal neutrons with specific probabilities to produce new isotopes whose mass number increases by one. These newly produced germanium isotopes are unstable and undergo spontaneous decay into other nuclides. 
These daughter nuclides serve as doping impurities within the germanium matrix, as shown in Table~\ref{tab:ntd_process}.
Germanium possesses an extremely low neutron capture cross section, hence the attenuation of the neutron flux within germanium samples can be neglected. This enables the NTD technique to achieve a highly uniform impurity distribution throughout the bulk of sample.
For high-purity germanium (HPGe) with natural isotopic abundance, the resulting NTD-Ge is a p-type semiconductor featuring complementary acceptor and donor doping with a fixed compensation ratio of $K=0.223$. 
The net doping concentration is given by,
\begin{equation}
   \centering
   n_{\rm{net}} = n({\rm ^{71}Ga}) - n({\rm ^{75}As})-2\cdot n({\rm ^{77}Se})= (1-K)\cdot n({\rm ^{71}Ga})\footnotemark.
   \label{eq:compensate_ration}
\end{equation}
\footnotetext{
The concentration $n({\rm ^{71}Ga})=n({\rm ^{71}Ge})=n_0\cdot \sigma\cdot \Phi$, where $n_0$ and $\sigma$ are the initial concentration and neutron capture cross section of ${\rm ^{70}Ge}$ respectively, $\Phi$ is the integrated neutron flux. The same relations apply to other impurities as well.} 

\begin{table}[hptb]
\centering
\small
\caption{Summaries of natural isotopic abundance~\cite{isotopicabundance} and thermal neutron capture cross-section $\sigma$~\cite{cross201889} for Ge isotopes, corresponding doping products and impurity types by NTD technique.}
\begin{tabularx}{0.6\textwidth}{ccccc}
\hline
Isotopes&Abundance~[\%]&Products& ~~~~~~$\sigma$~[b]~~~~~&Types\\
\hline
$^{70}{\rm Ge}$&20.52(19)&$^{71}{\rm Ga}$&3.05(13)&acceptor\\
$^{72}{\rm Ge}$&27.45(15)&$^{73}{\rm Ge}$&0.89(8)&-\\
$^{73}{\rm Ge}$&7.76(8)  &$^{74}{\rm Ge}$&14.7(4)&-\\
$^{74}{\rm Ge}$&36.52(12)&$^{75}{\rm As}$&0.36(4)&donor\\
$^{76}{\rm Ge}$&7.75(12) &$^{77}{\rm Se}$&0.055(2)&donor\\
\hline
\end{tabularx}
\label{tab:ntd_process}
\end{table}

The thermistor behavior of NTD-Ge stems from the hopping conduction mechanism. Below 1 K, thermal excitation energies are insufficient to ionize impurities, accordingly, acceptor and donor impurities cannot contribute mobile carriers within the valence and conduction bands.
Under such conditions, charge transport is governed by hopping conduction, which corresponds to direct electron tunneling between impurity sites in disordered compensated semiconductor systems~\cite{Efros_1975}. 
As phonon excitations grow with increasing temperature, electrons gain extra energy through electron–phonon interactions, and thus strengthen tunneling-mediated charge transport. Experiments have confirmed that the low-temperature resistivity ($\rho$) of NTD-Ge exhibits a temperature dependence well described by Mott’s law derived by Efros and Shklovskii~\cite{Efros_1975, MOTT19681} :
\begin{equation}
    \rho = \rho_0\cdot\exp{\sqrt{\frac{T_0}{T}}},
    \label{eq:ntd_resistive}
\end{equation}
where \(T_0\) is a critical parameter governing the sensitivity of NTD-Ge thermistors and strongly depends on doping concentration, whereas \(\rho_0\) is determined by multiple factors, predominantly the intrinsic properties of germanium substrates. 
Constrained by the metal-insulator transition (MIT) effect, Eq.~\ref{eq:ntd_resistive} is valid only for doping concentrations below the critical threshold. 
For uncompensated NTD-Ge fabricated from isotopically enriched \(^{70}\text{Ge}\), this critical doping concentration was reported as \(1.86\times10^{17}\ \text{cm}^{-3}\) by K. M. Itoh~\cite{PhysRevLett.77.4058}.

NTD-Ge thermistors exhibit high sensitivity to tiny temperature fluctuations under cryogenic conditions, making them ideal thermal sensors for cryogenic bolometer applications.
Cryogenic bolometers instrumented with NTD-Ge thermometers have demonstrated the outstanding detection performance, featuring a low energy threshold and excellent energy resolution~\cite{cuore_science, richochet_NTD, Guy:2024xog}.
In recent years, cryogenic bolometer technology has become increasingly essential for low-background fundamental physics research~\cite{Pretzl2020}. 
It underpins frontier investigations spanning searches for rare events including neutrinoless double beta decay (0$\nu\beta\beta$) and dark matter (DM), precision measurements of coherent elastic neutrino-nucleus scattering (CE$\nu$NS), as well as direct neutrino mass detection~\cite{cuore_science, richochet_NTD, nuleus_cevns, refId0}.
In China, a proposed 0$\nu\beta\beta$ search experiment employs Li$_2$$^{100}$MoO$_4$ scintillating cryogenic bolometer at the China Jinping Underground Laboratory (CJPL),
Within this experiment, NTD-Ge sensors will serve as the core readout devices for both heat and light signal channels~\cite{CJPL_bolometer, DUAN2026171161, CUPID_CJPL}.

Driven by the growing demand for high-performance cryogenic detectors and high-precision industrial temperature sensing, a research on the development of NTD-Ge thermometers has been carried out in China.
This paper elaborates on the detailed fabrication procedures and performance characterizations of NTD-Ge thermometers. The structure of this article is outlined below.
Section~\ref{sec:irradiation} describes the reactor-based neutron irradiation process adopted to produce NTD-Ge samples. 
Section~\ref{sec:defects} presents positron annihilation lifetime spectroscopy analyses for evaluating neutron irradiation-induced defects in NTD-Ge samples. 
Section~\ref{sec:hall} covers variable-temperature Hall measurements on NTD-Ge, which verify that the measured carrier concentration is consistent with the designed doping level. 
Lastly, prototype NTD-Ge thermometers are fabricated in Sec.~\ref{sec:low-temp}, and their low-temperature sensing performance is comprehensively validated.

\section{Reactor thermal neutron irradiation}\label{sec:irradiation}

Neutron irradiation experiments were carried out at the China Advanced Research Reactor (CARR), a multipurpose research reactor delivering a maximum thermal neutron flux of \(10^{15}\ \text{n}\cdot\text{cm}^{-2}\text{s}^{-1}\)~\cite{T40-2025-0538,CARR_2020}. 
The vertical irradiation channel employed in this work provides a neutron field with a thermal-to-fast neutron ratio greater than 99:1, which was calibrated using the cadmium-ratio method. A comprehensive overview of the CARR facility can be found in Ref.~\cite{CHEN2006966}.
The germanium wafers adopted in this study are 10N-grade high-purity, double-side-as-cut substrates provided by Umicore. 
The purity and natural isotopic abundance of these wafers were characterized using inductively coupled plasma mass spectrometry (ICP-MS) at the Shanghai Institute of Applied Physics. 
Table~\ref{tab:ge_isotope} presents the isotopic composition of the HP-Ge wafers measured via ICP-MS. 
The measurement agrees well with the natural isotopic abundance summarized in Table~\ref{tab:ntd_process}. 
The total native impurity fraction is lower than $10^{-10}$. 

\begin{table}[htbp]
\centering
\caption{ICP-MS-measured isotopic abundance of HP-Ge samples for NTD-Ge fabrication. }
\begin{tabularx}{0.55\textwidth}{cccccc}
\hline\hline
Isotopes &$^{70}$Ge& $^{72}$Ge& $^{73}$Ge & $^{74}$Ge& $^{76}$Ge \\
\hline
Abundance~[\%]&19.9&27.6&7.7&37.3&7.6\\
\hline\hline
\end{tabularx}\label{tab:ge_isotope}
\end{table}

\begin{figure}[!htbp]
	\centering
    \subfigure[]{\includegraphics[height=3.5cm]{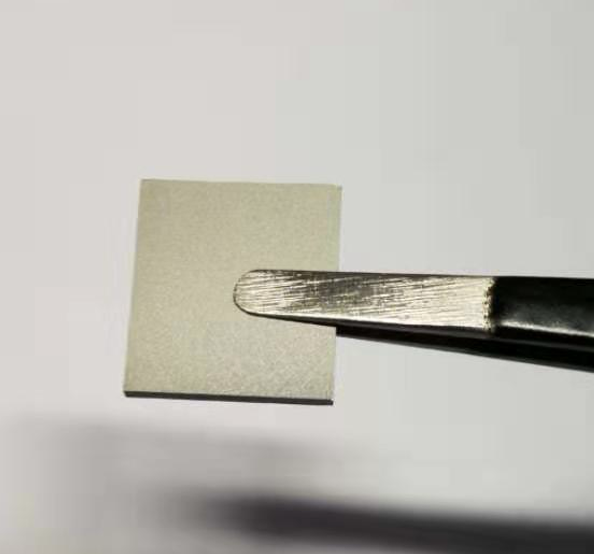}}
    \subfigure[]{\includegraphics[height=3.5cm]{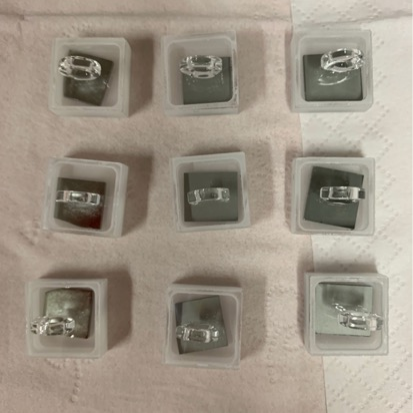}}
    \subfigure[]{\includegraphics[height=3.5cm]{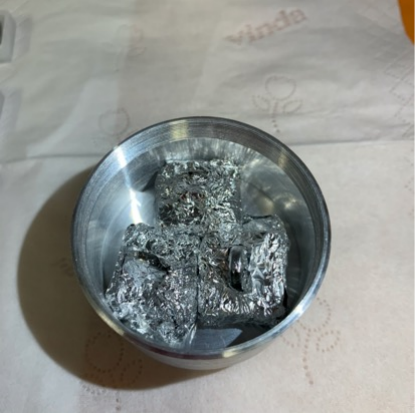}}
    \subfigure[]{\includegraphics[height=3.5cm]{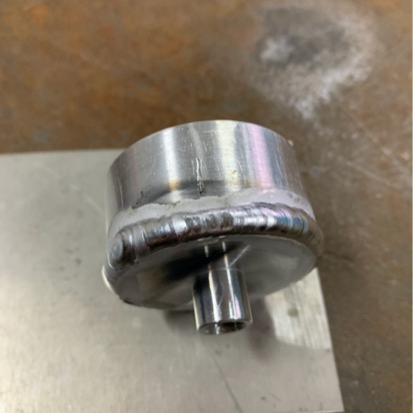}}
    \caption{HPGe sample treatment and nuclear reactor neutron irradiation preparation. (a) A piece of HPGe with size 10$\times$10$\times$1~mm$^3$. (b) The HPGe pieces are packaged in high purity thin quartz boxes. (c) Each quartz box is filled with high purity Al foil to prevent rocking. (d) Three quartz boxes form a set sealed in a high purity Al can.}
    \label{fig:Ge_sample}
\end{figure}

Prior to irradiation, pristine germanium wafers were diced into uniform sqaure pieces with standard dimensions of \(10\times10\times1\ \text{mm}^3\). In total, twenty-one HPGe samples were individually loaded into quartz containers and hermetically sealed within seven high-purity aluminum (Al) cans, with three samples encapsulated in each can. The macroscopic morphologies of the processed HPGe samples, quartz containers, and sealed aluminum cans are presented in Fig.~\ref{fig:Ge_sample}.

To fabricate NTD-Ge samples with doping concentrations spanning the MIT critical threshold, seven encapsulated aluminum cans were placed at discrete positions within the neutron irradiation field to achieve graded neutron fluence exposure. 
Among them, samples in two cans were irradiated under identical neutron fluence conditions, therefore, a total of six distinct neutron-irradiation fluence are expected. 
The entire neutron irradiation procedure was performed under stable reactor power conditions for a duration of 64.5 hours. 
The preliminary integrated neutron flux at each irradiation position was calculated using pre-calibrated neutron field parameters, as summarized in Table~\ref{tab:lowT_re}. 
To realize high-precision flux evaluation for individual HPGe samples, a novel self-monitoring technique was adopted, which relies on detecting characteristic X-rays emitted during the decay of in-situ produced \(^{71}{\rm Ge}\). 
During neutron irradiation, \(^{70}{\rm Ge}\) isotopes in HPGe samples capture thermal neutrons and undergo nuclear transmutation to \(^{71}{\rm Ge}\). 
The metastable \(^{71}{\rm Ge}\) subsequently decays into \(^{71}{\rm Ga}\), emitting characteristic \(K_\alpha\) (9.2 keV) and \(K_\beta\) (10.3 keV) X-rays. 
The integrated neutron flux can be quantitatively determined with high accuracy by measuring the X-ray yield. To eliminate interference from short-lived radionuclides, the neutron fluence measurements were conducted after a 150-day post-irradiation cooling period. 
The fundamental principles and detailed experimental procedures of this self-monitoring approach have been fully documented in Ref.~\cite{ZHAO2023168425}. The final calibrated integrated neutron flux values for all samples are listed in Table~\ref{tab:lowT_re}.

\begin{table}[width=.9\linewidth,cols=4,pos=h]
\caption{The pre-calibrated neutron fluence ($\Phi_{\rm pre}$) before irradiation, the measured neutron fluence ($\Phi_{\rm m}$) via self-monitoring method after irradiation, the corresponding Gallium concentration $n_{\rm Ga}$ 
and net doping concentration $n_{\rm net}$ of NTD-Ge samples, and the $T_0$ parameters determined by low-temperature measurements of fabricated thermometers.}\label{tab:lowT_re}
\begin{tabular*}{\tblwidth}{@{}CCCCCC@{} }
\toprule
\multirow{2}{*}{Thermometers} 
&$\Phi_{\rm pre}$&$\Phi_{\rm m}$&$n_{\rm Ga}$&$n_{\rm net}$&$T_0$ \\
&[$\times$10$^{18}$~n$\cdot$cm$^{-2}$]&[$\times$10$^{18}$~n$\cdot$cm$^{-2}$]&[$\times$10$^{17}$~cm$^{-3}$]&[$\times$10$^{17}$~cm$^{-3}$]&[K]\\
\midrule
NTD-Ge 7&2.01 &3.12 &0.86  &0.67  &11.72\\
NTD-Ge 6&3.02 &4.91 &1.36  &1.05  &8.45\\
NTD-Ge 5&3.52 &5.96 &1.64  &1.28  &4.30\\
NTD-Ge 4&3.83 &6.54 &1.80  &1.40  &3.97\\
NTD-Ge 3&4.33 &7.09 &1.96  &1.52  &Metal\\
NTD-Ge 2&5.03 &8.32 &2.30  &1.78  &Metal\\
\bottomrule
\end{tabular*}
\end{table}

\section{Defects characterization and thermal annealing}\label{sec:defects}
\begin{figure}[!htbp]
	\centering
    \includegraphics[height=6.0cm]{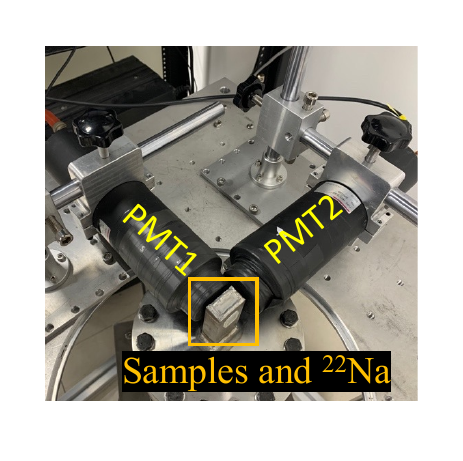}
    \includegraphics[height=6.0cm]{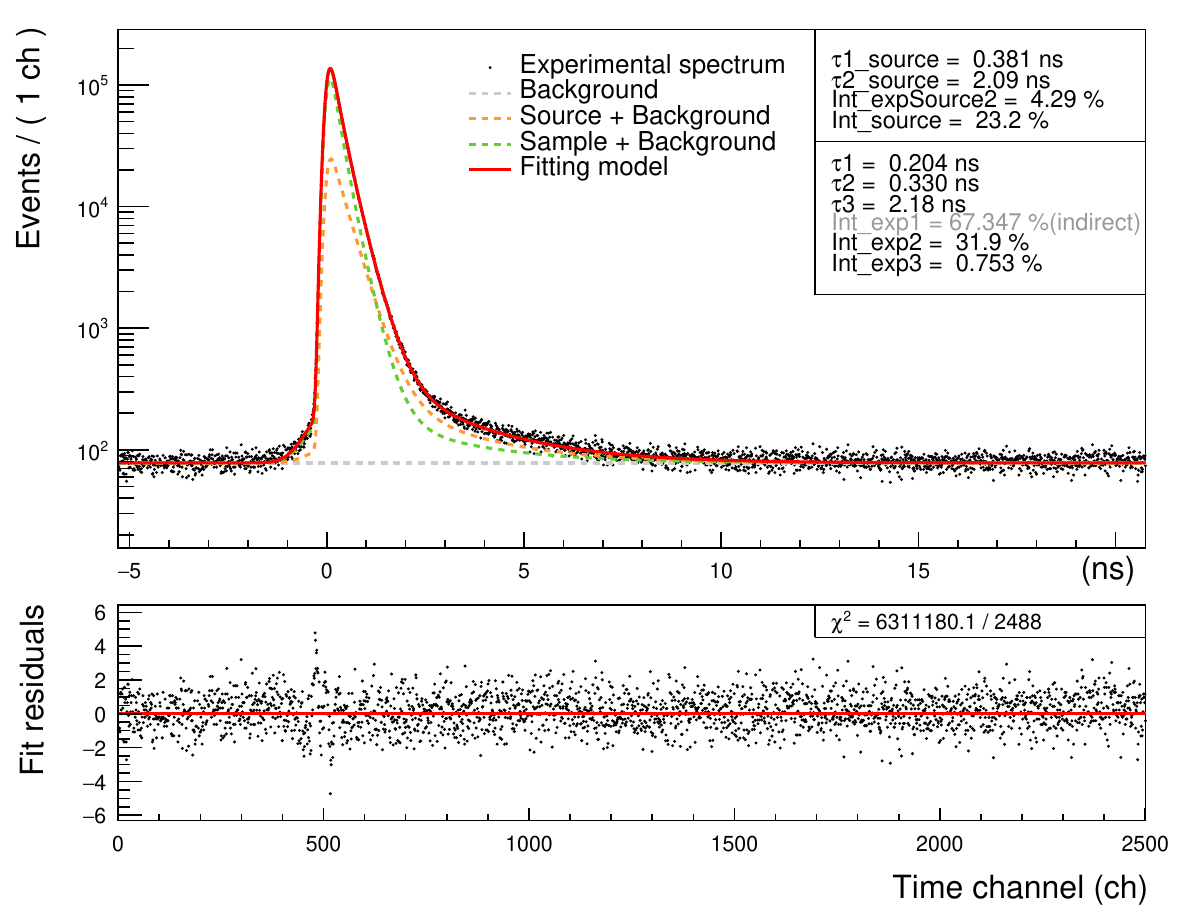}
    \caption{(left) Schematic of the PAL detector system, where he NTD-Ge sample and positron source are placed in vacuum environment during test;  (right) Representative experimental PAL spectrum of irradiated NTD-Ge sample, fitting with a function consists of the source component plus three exponential components from NTD-Ge contribution.
     }
    \label{fig:PALS}
\end{figure}

Neutron irradiation introduces radiation defects within germanium substrates.
For NTD-Ge samples, both bombardment by fast neutrons and nuclear recoils arising from radionuclide decay jointly generate substantial displacement defects. 
Such radiation damage creates mid-gap defect states and anomalous electrical properties, which can be eliminated through thermal annealing.
Palaio et al.~\cite{10.1063/1.333397} demonstrated that annealing at 400~$^\circ$C for 6 hours under an argon atmosphere can fully recover irradiation-induced damage. 
Accordingly, this annealing protocol was adopted throughout the present studies.
Positrons can be readily trapped by the microstructure features within materials and their lifetimes are highly sensitive to the local electron density.
This renders positron an ideal probe for defects including vacancies, voids, and dislocations. 
In this work, non-destructive positron annihilation lifetime (PAL) spectroscopy was performed at room temperature to characterize irradiation-induced defects in NTD-Ge samples~\cite{ZHAO2022166921}.
As illustrated in Fig.~\ref{fig:PALS}, a \(^{22}\)Na positron source (closed in two kapton foils) is sandwiched between two identical NTD-Ge slices from the same aluminum can. Two fast-response photomultiplier tubes (PMTs) were utilized to independently detect the two coincident photons, the energy 1.275~MeV photon associated with positron emission and the 0.511~MeV photon originating from electron-positron annihilation individually. 
The time interval of each paired coincident photons was recorded to derive the lifetime of individual positron annihilation event. 

In this measurement setup, the employed sample-source-sample sandwich structure ensures that positron annihilation exclusively within the NTD-Ge specimens or the Kapton foils of the positron source.
Consequently, the  acquired PAL spectra consist of two independent components, corresponding to positron annihilation in the NTD-Ge matrix (namely the NTD-Ge sample components) and the radioactive source (including the Kapton foils, namely the radioactive source component) respectively.
A representative PAL spectrum of the irradiated NTD-Ge is presented in Fig.~\ref{fig:PALS}.
Spectral fitting was performed by superimposing the individual spectral contributions from the NTD-Ge sample component and the radioactive source component.
The radioactive source component was modeled using two superimposed exponential function convoluted with a time resolution function.
All predefined source-related model parameters and component fractions were calibrated via reference PAL measurements on yttria-stabilized zirconia (YSZ), a standard material featuring well-defined and stable positron lifetime parameters, and subsequently were fixed throughout the fitting procedure for NTD-Ge spectra.
The NTD-Ge sample component was decomposed into three exponential terms. 
The first two terms yield two distinct lifetime values ($\tau_1$ and $\tau_2$) corresponding to the two-state trapping model, while the third term with negligible intensity was incorporated to optimize the overall fitting quality.
Specifically, $\tau_2$ characterizes the positron annihilation lifetime at defects sites within NTD-Ge bulk, while the short-lived component, $\tau_1=1/(\lambda_b+\kappa_d)$, is shorter than the bulk positron lifetime $\lambda_b$, which characterizes the positron annihilation lifetime in defect-free germanium lattice. Here $\kappa_d$ refers to the positron trapping rate of lattice defects.

The fitted positron lifetimes \(\tau_1\) and \(\tau_2\), as well as the corresponding intensities of NTD-Ge samples irradiated under different neutron fluences are summarized in Table~\ref{tab:pals_data}. 
The \(\tau_2\) values from 310 to 330~ps exhibit nearly independence on neutron fluence.
This lifetime feature is highly consistent with the typical signature of divacancy defects in germanium, verifying that neutron irradiation predominantly generates divacancy defects~\cite{PhysRevB.83.235212}. 
In contrast, the intensity \(I_2\), corresponding to the \(\tau_2\) component, increases significantly with increasing neutron fluence, demonstrating that the density of irradiation-induced divacancy defects rises with elevation of neutron fluence.
Thermal annealing was implemented to substantially eliminate irradiation-induced defects in NTD-Ge. 
The PAL results for two heavily irradiated NTD-Ge samples after thermal annealing are also listed in Table~\ref{tab:pals_data}. 
After annealing, the \(\tau_2\) value of the NTD-Ge spectral component rises to above 400~ps, while its corresponding intensity drops dramatically to nearly zero. 
Meanwhile, the \(\tau_1\) value approaches 228~ps, which is the bulk positron lifetime \(\tau_b\) of a defect-free germanium lattice.
All these phenomena can be attributed to the migration and aggregation of irradiation-induced vacancies during thermal annealing, which promotes the formation of vacancy clusters and enlarged divacancy structures. 
These experimental observations distinctly demonstrate that neutron irradiation introduces a high density of isolated divacancies in NTD-Ge, while post-irradiation annealing effectively eliminates these primary defects and potentially facilitates the formation of a small number of large-scale vacancy complexes.

\begin{table}[width=.9\linewidth,cols=4,pos=h]
\caption{Summaries of fitting on measured PAL spectra for NTD-Ge with the different irradiated neutron fluences with and without thermal annealing.}\label{tab:pals_data}
\begin{tabular*}{\tblwidth}{@{}LCCCCR@{} }
\toprule
\multirow{2}{*}{Samples} 
&Neutron fluence&$\tau_1$ &$I_1$ &$\tau_2$ &$I_2$\\
&[$\times$10$^{18}$~n$\cdot$cm$^{-2}$]&[ps]&[\%]&[ps]&[\%]\\
\midrule
NTD-Ge~7&3.21 &203.3$\pm$3.6  &66.7$\pm$2.7   &330.5$\pm$8.8   &32.5$\pm$2.7\\
NTD-Ge~6&5.05 &176.8$\pm$2.4  &50.5$\pm$1.1   &316.7$\pm$2.6   &47.8$\pm$1.1\\
NTD-Ge~5&5.80 &182.9$\pm$2.7  &49.2$\pm$1.5   &313.5$\pm$0.7   &50.0$\pm$1.5\\
NTD-Ge~4&6.41 &181.3$\pm$0.5  &42.9$\pm$0.2   &313.5$\pm$0.7   &56.8$\pm$0.2\\
NTD-Ge~3&7.36 &158.3$\pm$2.9  &39.7$\pm$1.1   &311.0$\pm$2.8   &58.5$\pm$1.1\\
NTD-Ge~2&8.07 &167.6$\pm$0.4  &33.7$\pm$0.3   &312.3$\pm$0.8   &65.7$\pm$0.3\\ \hline
NTD-Ge~2 (annealed)&8.07 &\textbf{227.8$\pm$1.9} &\textbf{95.5$\pm$2.1 }&402.0$\pm$61 &4.3$\pm$2.1\\
NTD-Ge~3 (annealed)&7.36 &\textbf{228.0$\pm$0.1} &\textbf{95.1$\pm$0.1} &605.0$\pm$30  &2.9$\pm$0.1\\
\bottomrule
\end{tabular*}
\end{table}

\section{Hall measurements at variable temperatures}\label{sec:hall}

Hall effect measurements were carried out to electrically quantify the doping level of NTD-Ge samples, whose magnitude is directly governed by the neutron fluence applied during irradiation.
For semiconductors materials, Hall characterization enables the quantitative determination of its mobile carrier density.
When mobile carriers originate predominantly from the ionization of doping impurities under full ionization conditions, the measured carrier density is equivalent to the net impurity concentration of the material.
The experimentally derived Hall coefficient is determined as
\begin{equation}
R_H=\frac{tV_H}{BI}=r_H\cdot\frac{1}{nq},
\end{equation}
where $V_H$ is the Hall voltage, $I$ is the applied current, $B$ represents the applied magnetic field and $t$ denotes the sample thickness. 
The second formula expresses the theoretical correlation between the Hall coefficient $R_H$ and mobile carriers density $n$, where $r_H$ is the Hall factor and $q$ is the elementary charge. 
As a constant governed by carrier scattering mechanisms, the Hall factor \(r_H\) generally deviates from unity. 
For compensated p-type NTD-Ge samples, the value of $r_H$ ranges from 0.8 to 1.9~\cite{li2012semiconductor}. 
\begin{figure}[!h]
\centering
\includegraphics[height=5.5cm]{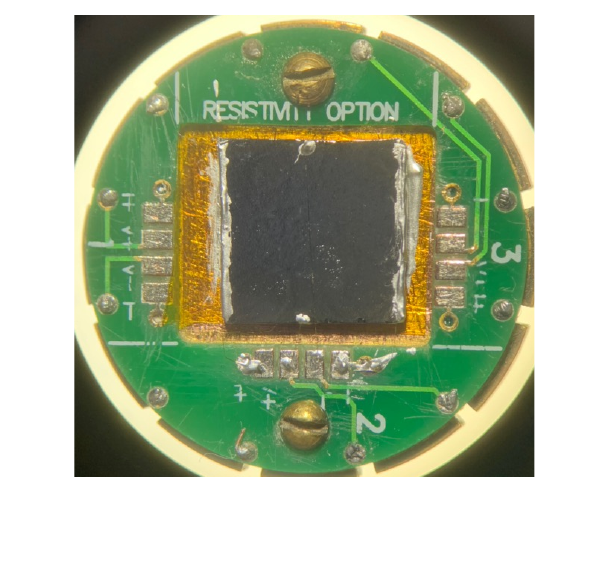}
\includegraphics[height=6.0cm]{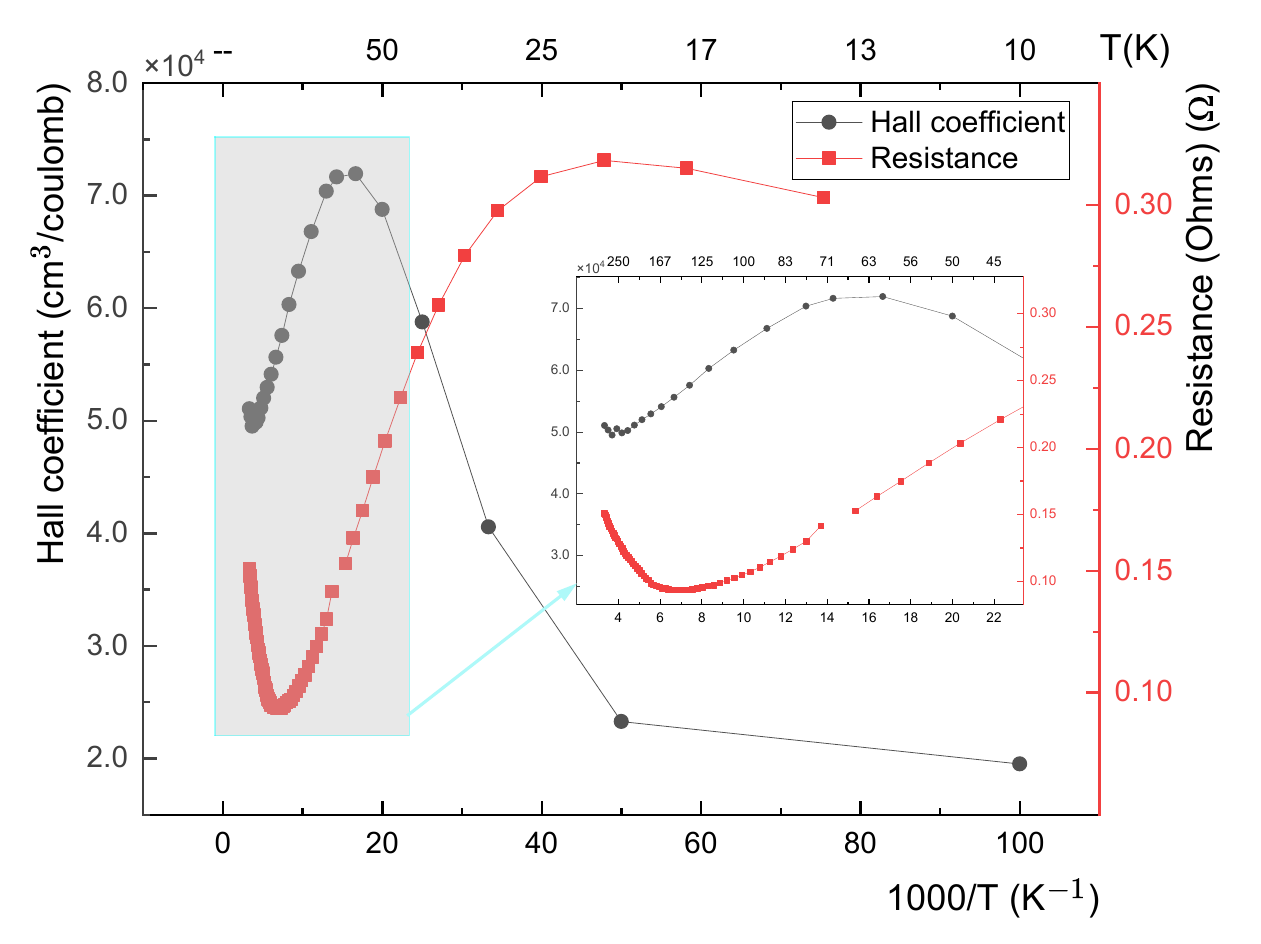}
\caption{(left)~Photograph of a NTD-Ge sample prepared for Hall effect measurements. (right)~Temperature-dependent Hall coefficient and resistance results of a NTD-Ge SAMPLE measured by PPMS.}
\label{fig:Hall_measurement}
\end{figure}

Investigating the temperature-dependent characteristics of the Hall coefficient is crucial for verifying the complete ionization of impurity atoms in NTD-Ge. 
Accordingly, temperature-variable Hall effect measurements were conducted from 300~K down to 10~K using a Physical Property Measurement System (PPMS). 
As shown in the left panel of Fig.~\ref{fig:Hall_measurement}, electrical contacts for Hall testing were fabricated in a Van der Pauw configuration using silver epoxy and 25~${\rm{\mu}}$m diameter gold wires. 
The silver epoxy electrodes exhibited reliable ohmic contact throughout the entire measured temperature range and could be easily stripped off with acetone to restore the original sample surface. 
To eliminate measurement uncertainties originating from asymmetric electrode geometry, Hall voltages were acquired under magnetic fields sweeping from -1.5~T to +1.5~T.

\begin{figure}[!htbp]
\centering
\includegraphics[height=6.0cm]{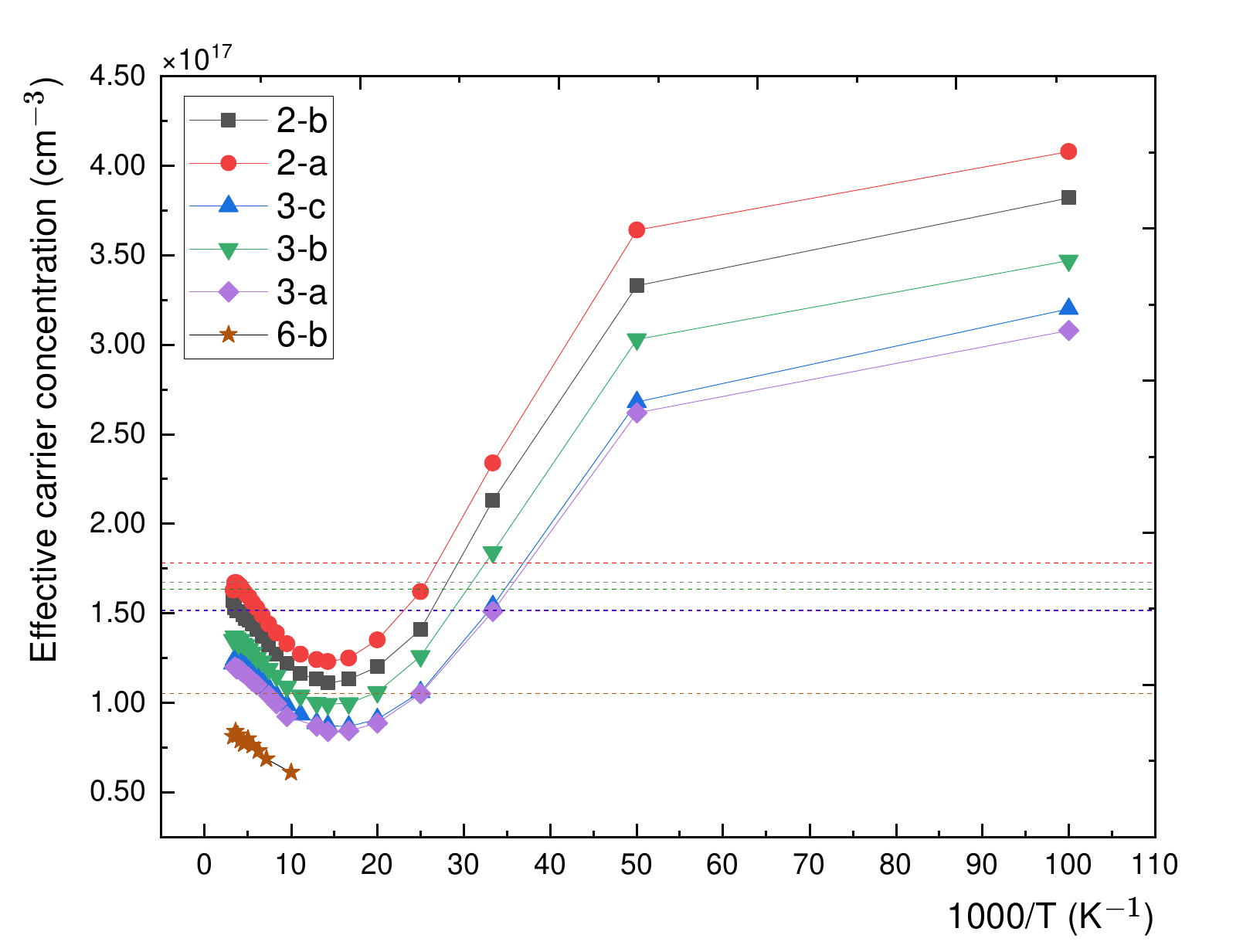}
\caption{The effective carrier concentration with the assumption of $r_H=1$ as a function of $1000/T$, where a, b, c denote different pieces in the same Al container during irradiation. The dashed lines indicate the net acceptor doping concentration corresponding to the neutron fluence.}
\label{fig:Effective_carrier}
\end{figure}

The right panel of Fig.~\ref{fig:Hall_measurement} illustrates the temperature-dependence evolution of the Hall coefficient and two-electrode resistance of a NTD-Ge sample.
In the temperature range from 300~K to 250~K, the Hall coefficient remains low and nearly constant, which can be conservatively attribute to the saturated ionization state of dopant impurities. 
As the temperature decreases to approximately 150~K, the electrical resistance of the sample decreases, accompanied by a slight rise in the Hall coefficient. 
This phenomenon is attributed to the weakened phonon scattering at lower temperatures, which leads to an increase in carrier mobility~\cite{PhysRev.119.1238,Shklovskii1984}. 
Further cooling below $\sim$150~K induces gradual impurity freeze-out, which reduces the mobile carrier density and leads to a substantial increase in both electrical resistance and Hall coefficient.
This result verifies that impurity freeze-out initiates at $\sim$150~K and persists at lower temperatures for the tested NTD-Ge samples. 
A distinct maximum value appears on the Hall coefficient curve at $\sim$77~K, revealing a transition of the dominant conduction mechanisms from band conduction to impurity hopping conduction. 
Above this transition temperature, electrical transport is predominantly governed by valence-band carriers generated from fully ionized doping impurities. 
As cooling continues, massive impurities fall into the freeze-out state, and impurity-mediated hopping conduction gradually dominates the electrical conduction.
This process equivalently increases the effective carrier density and modulates the overall electrical transport properties of the sample.
Accordingly, impurity ionization saturation can be reliably confirmed at approximately 300~K. 
With the assumption of a unity Hall factor $r_H=1$, the effective mobile carrier density of the measured NTD-Ge samples is shown in Fig.~\ref{fig:Effective_carrier}. 
Due to the absence of precise values of Hall factor, absolute carrier concentrations cannot be precisely quantified from Hall measurements. 
Nevertheless, the Hall measurement results strongly verify the full ionization and activation of dopant impurities. Furthermore, the impurity concentrations derived under the impurity ionization saturation state exhibit good consistency with the doping levels calibrated from neutron fluences using the self-monitoring method~\cite{ZHAO2023168425}.

\section{Fabrication and cryogenic performance characterization of NTD-Ge thermometers}\label{sec:low-temp}
\begin{figure}[!htbp]
    \centering
    \includegraphics[height=5.0cm]{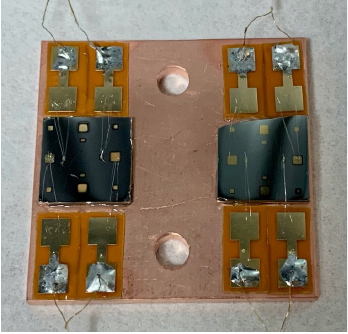}
    \caption{Photograph of the fabricated NTD-Ge thermometer.}\label{fig:NTD}
\end{figure}

NTD-Ge thermometers require the fabrication of surface Ohmic contacts to establish stable and robust electrical connection. 
A standard and widely adopted approach is to form a thin and heavily doped surface layer with the same doping type as the semiconductor bulk. 
This strategy can effectively narrow the width of Schottky barrier between the metal electrode and semiconductor, enabling freely bidirectional carrier transport through quantum tunneling.
Prior to device fabrication, NTD-Ge samples were thermally annealed to eliminate lattice defects induced by neutron irradiation.
Subsequent chemical mechanical polishing (CMP) is performed to remove the native oxide layer and yield an atomic-scale flat surface, which was verified by atomic force microscopy (AFM) characterization. 
Boron ion implantation was then conducted to fabricate a p++ heavily doped layer on the surface of p-type NTD-Ge samples. A two-step implantation process is adopted with sequential parameters: a dose of \(2\times10^{14}\ \mathrm{ions/cm^2}\) at an acceleration energy of 50~keV, followed by a dose of \(1\times10^{14}\ \mathrm{ions/cm^2}\) at 33 keV. 
According to SRIM simulation~\cite{ZIEGLER20101818}, this implantation scheme can produce a uniformly doped layer with a depth of 300~nm with a doping concentration of approximately \(10^{19}\ \mathrm{cm^{-3}}\). 
After boron ion implantation, composite metallic electrodes consisting of a 20~nm nickel adhesion layer and a 400~nm gold conductive layer were deposited on the implanted regions via magnetron sputtering. Notably, argon ion bombardment pretreatment before metal deposition is essential to ensure strong and robust adhesion between metallic films and the Ge substrate.  
In the end, the fabricated NTD-Ge thermometers were post-annealed at 250 \(^\circ\mathrm{C}\) for 1~h in a nitrogen atmosphere to repair lattice defects generated during the ion implantation process. For the purposes of structural redundancy and improved device reliability, multiple electrodes are patterned on the same surface of the NTD-Ge samples, as illustrated in Fig.~\ref{fig:NTD}.

\begin{figure}[!htbp]
\centering
\includegraphics[height=6.0cm]{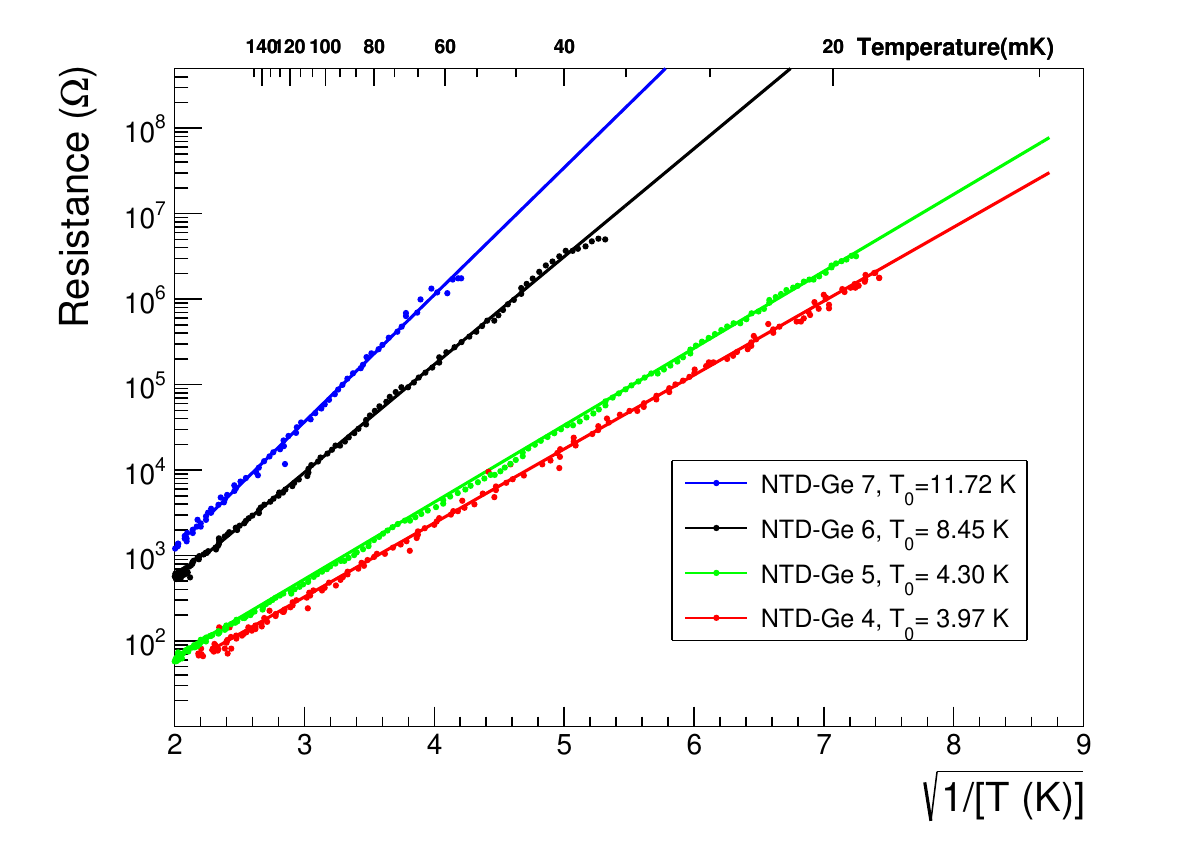}
\caption{Temperature dependence of resistances of NTD-Ge thermometers at $T<0.25~{\rm K}$.\label{fig:LT_measurement}}
\end{figure}

Cryogenic performance characterization of the fabricated NTD-Ge thermometers was carried out on the Cryogenic Detector R\&D Platform at the University of Science and Technology of China (USTC). 
All NTD-Ge thermometers were affixed to a copper holder using Araldite\textregistered epoxy to establish robust thermal coupling to the 10-mK stage of the cryostat. 
Each electrode was wire-bonded to a Kapton transfer pad using gold wires, subsequently connected to the internal cryostat socket via a soldered NbTi superconducting wire.
Throughout the cryogenic measurements, the temperature of the mixing chamber plate was continuously monitored using a calibrated RuO$_{2}$ thermometer. 
Two independent methods were employed to measure the resistance of NTD-Ge thermometers at each set temperature. 
The first method makes use of a custom-built electronic system integrating a bias circuit and a differential preamplifier, enabling precise acquisition of current–voltage (I–V) curves at each target temperature. 
The second one utilizes a commercial AC resistance bridge (Lakeshore Model 372) to achieve rapid and direct resistance readouts, allowing for automated continuous measurements during a free cooldown of the cryostat.
The measured I–V curves exhibit excellent linearity and polarity symmetry, verifying the reliable ohmic behavior of the fabricated contacts. 
The linear I–V characteristics indicate negligible Joule heating  induced during electrical testing. 
The resistance values extracted from I-V fitting are in excellent agreement with those obtained by the Lakeshore system for the same channel across the entire test temperature range. 
Figure~\ref{fig:LT_measurement} illustrates the temperature-dependent resistance characteristics of four NTD-Ge thermometers with distinct doping concentrations. 
The others two samples with higher doping concentrations exhibit ultralow resistance and metal-like electrical characteristics at cryogenic temperatures, and thus the corresponding data are not included herein.
The temperature-resistance curves of the valid samples are well fitted by Eq.~\ref{eq:ntd_resistive}, and the corresponding fitting results are displayed in Fig.~\ref{fig:LT_measurement}. 
All \(T_0\) parameters extracted from the fitting are summarized in Table~\ref{tab:lowT_re}.


As shown in Fig.~\ref{fig:LT_measurement}, the temperature-dependent resistance of the four functional NTD‑Ge thermometer follows Mott's law down to approximately $\sim$20~mK. 
The resistive characteristics measured in the mK regime validate the  working principle of NTD-Ge thermometer, demonstrating the feasibility and reliability of the entire fabrication process. 
The \(T_0\) values derived from the Mott-law fitting cover the optimal parameter range for bolometric applications in the cutting-edge fundamental research $0\nu\beta\beta$ experiments, referring to $\sim$4~K adopted in the CUORE/CUPID experiment~\cite{Alfonso_2023}. 
Furthermore, resistance measurements acquired across different electrode pairs of the same device yield highly consistent \(T_0\) values. 
Such reproducibility verifies that \(T_0\) parameter is predominantly governed by the doping concentration, which is full consistent with the fundamental theory of variable-range hopping. 

By contrast, the two NTD-Ge thermometers irradiated at high neutron fluence exhibit metal-like electrical properties at cryogenic temperatures.
This phenomenon originates from the doping concentrations exceeding the MIT critical concentration, as introduced in Sec.~\ref{sec::intro}. 
This observation aligns with the established fact that the MIT critical Gallium concentration is higher for compensated NTD‑Ge samples relative to their uncompensated counterparts. 
This phenomenon stems from multiple physical effects present in compensated systems, including broadened impurity bandwidth induced by enhanced carrier scattering and  weakened screening of impurity Coulomb potentials, as elaborated in previous literature~\cite{Fritzsche01121980}. 

\section{Conclusion and prospects}
This work presents a systematic study of NTD-Ge thermometers for applications in cryogenic bolometric detector. 
A series of HPGe samples were neutron-irradiated at CARR to fabricate compensated p-type NTD-Ge materials with well controlled doping levels. 
Positron annihilation lifetime measurements were performed to characterize irradiation-induced lattice defects in NTD-Ge samples, and the results confirm that the applied thermal annealing can effectively eliminate radiation defects. 
Meanwhile, variable-temperature Hall measurements were carried out to investigate the temperature-dependent carrier transport characteristics. 
The extracted impurity concentrations exhibit good consistency with the values evaluated from the neutron irradiation fluence.
High-performance NTD-Ge thermometers were successfully fabricated through a sequential process involving surface polishing, boron ion implantation, and Ni/Au metallization, yielding stable and high-quality Ohmic contacts for the devices.
Cryogenic electrical‑characterization measurements down to the mK regime  verify that the temperature-dependent resistance of the fabricated thermometers closely follows the Efros–Shklovskii variable-range hopping law,  with the extracted $T_0$ parameters well consistent with expectations. 
This study establishes a comprehensive and indigenous fabrication route for high-quality NTD-Ge thermometers and systematically validates their low-temperature resistive properties. 
The established correlation between $T_0$ and doping concentration provides a fundamental basis for future neutron irradiation optimization and performance investigation of NTD-Ge thermometers toward diverse application for various scenarios. 
Future efforts will be devoted to optimizing the fabrication specifications of NTD-Ge thermometers and carrying out in-situ performance validation by integrating these thermometers into bolometric  detector system.

\section{Acknowledgment}
This work is supported by National Key R\&D Program of China (2023YFA1607203), National Natural Science Foundation of China (12475190, W2543016), the Fundamental Research Funds for the Central Universities, China (WK2030250141). This work was partially performed at the University of Science and Technology of China (USTC) Center for Micro- and Nanoscale Research and Fabrication, and we thank Yu Wei and Haitao Liu for their kind assistance in the processing and fabrication of samples.

\bibliographystyle{model1-num-names}

\bibliography{biblio}

\end{document}